\pdfoutput=1

\documentclass[12pt,a4paper]{article}

\usepackage{ifthen} 
\newboolean{pdflatex}
\setboolean{pdflatex}{true} 

\newboolean{articletitles}
\setboolean{articletitles}{true} 

\newboolean{uprightparticles}
\setboolean{uprightparticles}{false} 

\newboolean{inbibliography}
\setboolean{inbibliography}{false} 

\usepackage{microtype}
\usepackage{lineno}  
\usepackage{xspace} 
\usepackage{caption} 

\usepackage{graphicx}  
\usepackage{color}
\usepackage{colortbl}
\graphicspath{{./figs/}} 

\usepackage{amsmath} 
\usepackage{amssymb}
\usepackage{amsfonts}
\usepackage{upgreek} 

\newcommand*\patchAmsMathEnvironmentForLineno[1]{%
\expandafter\let\csname old#1\expandafter\endcsname\csname #1\endcsname
\expandafter\let\csname oldend#1\expandafter\endcsname\csname
end#1\endcsname
 \renewenvironment{#1}%
   {\linenomath\csname old#1\endcsname}%
   {\csname oldend#1\endcsname\endlinenomath}%
}
\newcommand*\patchBothAmsMathEnvironmentsForLineno[1]{%
  \patchAmsMathEnvironmentForLineno{#1}%
  \patchAmsMathEnvironmentForLineno{#1*}%
}
\AtBeginDocument{%
\patchBothAmsMathEnvironmentsForLineno{equation}%
\patchBothAmsMathEnvironmentsForLineno{align}%
\patchBothAmsMathEnvironmentsForLineno{flalign}%
\patchBothAmsMathEnvironmentsForLineno{alignat}%
\patchBothAmsMathEnvironmentsForLineno{gather}%
\patchBothAmsMathEnvironmentsForLineno{multline}%
\patchBothAmsMathEnvironmentsForLineno{eqnarray}%
}

\usepackage{hyperref}    
\usepackage[all]{hypcap} 

\def\lhcb {\mbox{LHCb}\xspace}

\def\babar  {\mbox{BaBar}\xspace}
\def\belle  {\mbox{Belle}\xspace}

\def\cern {\mbox{CERN}\xspace}

\def\MagUp {\mbox{\em Mag\kern -0.05em Up}\xspace}

\ifthenelse{\boolean{uprightparticles}}%
{

 \def\Ppi         {\ensuremath{\uppi}\xspace}                 
                  
 \def\Prho        {\ensuremath{\uprho}\xspace}

 \def\Ppsi        {\ensuremath{\uppsi}\xspace}

 \def\PDelta      {\ensuremath{\Delta}\xspace}                 
 \def\PXi      {\ensuremath{\Xi}\xspace}                 
 \def\PLambda      {\ensuremath{\Lambda}\xspace}                 
 \def\PSigma      {\ensuremath{\Sigma}\xspace}                 
 \def\POmega      {\ensuremath{\Omega}\xspace}                 
 \def\PUpsilon      {\ensuremath{\Upsilon}\xspace}                 
 
 \def\PB      {\ensuremath{\mathrm{B}}\xspace}                 
                  
 \def\PD      {\ensuremath{\mathrm{D}}\xspace}

 \def\PJ      {\ensuremath{\mathrm{J}}\xspace}                 
 \def\PK      {\ensuremath{\mathrm{K}}\xspace}

 \def\Pb      {\ensuremath{\mathrm{b}}\xspace}                 
 \def\Pc      {\ensuremath{\mathrm{c}}\xspace}

 \def\Pi      {\ensuremath{\mathrm{i}}\xspace}

 \def\Ps      {\ensuremath{\mathrm{s}}\xspace}

}
{

 \def\Ppi         {\ensuremath{\pi}\xspace}                 
                  
 \def\Prho        {\ensuremath{\rho}\xspace}

 \def\Ppsi        {\ensuremath{\psi}\xspace}                 
                  
 \mathchardef\PDelta="7101
 \mathchardef\PXi="7104
 \mathchardef\PLambda="7103
 \mathchardef\PSigma="7106
 \mathchardef\POmega="710A
 \mathchardef\PUpsilon="7107
                  
 \def\PB      {\ensuremath{B}\xspace}                 
                  
 \def\PD      {\ensuremath{D}\xspace}

 \def\PJ      {\ensuremath{J}\xspace}                 
 \def\PK      {\ensuremath{K}\xspace}

 \def\Pb      {\ensuremath{b}\xspace}                 
 \def\Pc      {\ensuremath{c}\xspace}

 \def\Pi      {\ensuremath{i}\xspace}

 \def\Ps      {\ensuremath{s}\xspace}

}

\makeatletter
\ifcase \@ptsize \relax
  \newcommand{\miniscule}{\@setfontsize\miniscule{4}{5}}
\or
  \newcommand{\miniscule}{\@setfontsize\miniscule{5}{6}}
\or
  \newcommand{\miniscule}{\@setfontsize\miniscule{5}{6}}
\fi
\makeatother

\DeclareRobustCommand{\optbar}[1]{\shortstack{{\miniscule (\rule[.5ex]{1.25em}{.18mm})}
  \\ [-.7ex] $#1$}}

\def\squark    {{\ensuremath{\Ps}}\xspace}

\def\cquark    {{\ensuremath{\Pc}}\xspace}

\def\bquark    {{\ensuremath{\Pb}}\xspace}

\def\pion   {{\ensuremath{\Ppi}}\xspace}

\def\rhoz     {{\ensuremath{\rhomeson^0}}\xspace}

\def\kaon    {{\ensuremath{\PK}}\xspace}
  \def\Kbar    {{\kern 0.2em\overline{\kern -0.2em \PK}{}}\xspace}

\def\KorKbar    {\kern 0.18em\optbar{\kern -0.18em K}{}\xspace}

\def\Kstarz  {{\ensuremath{\kaon^{*}(892)^{0}}}\xspace}

  \def\Dbar    {{\kern 0.2em\overline{\kern -0.2em \PD}{}}\xspace}
\def\D       {{\ensuremath{\PD}}\xspace}

\def\DorDbar    {\kern 0.18em\optbar{\kern -0.18em D}{}\xspace}
\def\Dz      {{\ensuremath{\D^0}}\xspace}

\def\B       {{\ensuremath{\PB}}\xspace}
\def\Bbar    {{\ensuremath{\kern 0.18em\overline{\kern -0.18em \PB}{}}}\xspace}

\def\BorBbar    {\kern 0.18em\optbar{\kern -0.18em B}{}\xspace}
\def\Bz      {{\ensuremath{\B^0}}\xspace}

\def\Bu      {{\ensuremath{\B^+}}\xspace}

\def\Bp      {{\ensuremath{\Bu}}\xspace}

\def\Bd      {{\ensuremath{\B^0}}\xspace}
\def\Bs      {{\ensuremath{\B^0_\squark}}\xspace}

\def\Bdb     {{\ensuremath{\Bbar{}^0}}\xspace}

\def\jpsi     {{\ensuremath{{\PJ\mskip -3mu/\mskip -2mu\Ppsi\mskip 2mu}}}\xspace}

  \def\Y#1S{\ensuremath{\PUpsilon{(#1S)}}\xspace}

\def\Lbar        {{\ensuremath{\kern 0.1em\overline{\kern -0.1em\PLambda}}}\xspace}
\def\LorLbar    {\kern 0.18em\optbar{\kern -0.18em \PLambda}{}\xspace}

\def\BF         {{\ensuremath{\cal B}}\xspace}

\def\BR         {\BF}
\newcommand{\decay}[2]{\ensuremath{#1\!\to #2}\xspace}         

\def\to                 {\ensuremath{\rightarrow}\xspace}

\def\CP                {{\ensuremath{C\!P}}\xspace}

\def\AT#1     {\ensuremath{A_{\mathrm{T}}^{#1}}\xspace}           

\def\C#1      {\ensuremath{\mathcal{C}_{#1}}\xspace}                       
\def\Cp#1     {\ensuremath{\mathcal{C}_{#1}^{'}}\xspace}                    
\def\Ceff#1   {\ensuremath{\mathcal{C}_{#1}^{\mathrm{(eff)}}}\xspace}        
\def\Cpeff#1  {\ensuremath{\mathcal{C}_{#1}^{'\mathrm{(eff)}}}\xspace}       
\def\Ope#1    {\ensuremath{\mathcal{O}_{#1}}\xspace}                       
\def\Opep#1   {\ensuremath{\mathcal{O}_{#1}^{'}}\xspace}                    

\newcommand{\tev}{\ifthenelse{\boolean{inbibliography}}{\ensuremath{~T\kern -0.05em eV}\xspace}{\ensuremath{\mathrm{\,Te\kern -0.1em V}}}\xspace}
\newcommand{\gev}{\ensuremath{\mathrm{\,Ge\kern -0.1em V}}\xspace}
\newcommand{\mev}{\ensuremath{\mathrm{\,Me\kern -0.1em V}}\xspace}
\newcommand{\kev}{\ensuremath{\mathrm{\,ke\kern -0.1em V}}\xspace}
\newcommand{\ev}{\ensuremath{\mathrm{\,e\kern -0.1em V}}\xspace}
\newcommand{\gevc}{\ensuremath{{\mathrm{\,Ge\kern -0.1em V\!/}c}}\xspace}
\newcommand{\mevc}{\ensuremath{{\mathrm{\,Me\kern -0.1em V\!/}c}}\xspace}
\newcommand{\gevcc}{\ensuremath{{\mathrm{\,Ge\kern -0.1em V\!/}c^2}}\xspace}
\newcommand{\gevgevcccc}{\ensuremath{{\mathrm{\,Ge\kern -0.1em V^2\!/}c^4}}\xspace}
\newcommand{\mevcc}{\ensuremath{{\mathrm{\,Me\kern -0.1em V\!/}c^2}}\xspace}

\def\mum  {\ensuremath{{\,\upmu\rm m}}\xspace}

\def\invfb   {\ensuremath{\mbox{\,fb}^{-1}}\xspace}

\newcommand{\stat}{\ensuremath{\mathrm{\,(stat)}}\xspace}
\newcommand{\syst}{\ensuremath{\mathrm{\,(syst)}}\xspace}

\def\deriv {\ensuremath{\mathrm{d}}}

\def\gsim{{~\raise.15em\hbox{$>$}\kern-.85em
          \lower.35em\hbox{$\sim$}~}\xspace}
\def\lsim{{~\raise.15em\hbox{$<$}\kern-.85em
          \lower.35em\hbox{$\sim$}~}\xspace}

\newcommand{\Real}{\ensuremath{\mathcal{R}e}\xspace}

\def\sPlot{\mbox{\em sPlot}\xspace}

\def\ptot       {\mbox{$p$}\xspace}
\def\pt         {\mbox{$p_{\rm T}$}\xspace}

\def\evtgen     {\mbox{\textsc{EvtGen}}\xspace}

\def\geant      {\mbox{\textsc{Geant4}}\xspace}

\def\pythia     {\mbox{\textsc{Pythia}}\xspace}

\def\tell1  {TELL1\xspace}
\def\ukl1   {UKL1\xspace}

\usepackage{cite} 
\usepackage{mciteplus}

\usepackage{longtable} 

\ifthenelse{\boolean{uprightparticles}}%
{\def\Prho      {\ensuremath{\uprho}\xspace}
 
}
{\def\Prho      {\ensuremath{\rho}\xspace}
 
}
\def\rhoz   {\ensuremath{\Prho^0}\xspace}

\def\KPi {\ensuremath{(\kaon^+\pion^-)}\xspace}
\def\PiPi {\ensuremath{(\pion^+\pion^-)}\xspace}
\def\KK  {\ensuremath{(\kaon^+\kaon^-)}\xspace}

\def\BtoRhoRho  {\decay{\B}{\rho^0\rho^0}}
\def\BdtoRhoRho  {\decay{\Bd}{\rho^0\rho^0}}

\def\BdtoPhiKst  {\decay{\Bd}{\phi\Kstarz}}

\def\PiPiPiPi {\ensuremath{(\pion^+\pion^-)(\pion^+\pion^-)}\xspace}

\def\BtoPiPiPiPi   {\ensuremath{{\decay{\Bd}{(\pion^+\pion^-)(\pion^+\pion^-)}}}\xspace}

\def\BstoPiPiPiPi  {\decay{\Bs}{(\pion^+\pion^-)(\pion^+\pion^-)}\xspace}

\def\BtoKPiPiPi  {\decay{\Bd}{(\kaon^+\pion^-)(\pion^+\pion^-)}\xspace}

\def\BstoKPiPiPi  {\decay{\Bs}{(\kaon^-\pion^+)(\pion^+\pion^-)}\xspace}

\def\BtoKKKPi  {\decay{\Bd}{(\kaon^+\kaon^-)(\kaon^{+}\pion^{-})}\xspace}

\def\BstoKKKPi  {\decay{\Bs}{(\kaon^+\kaon^-)(\kaon^-\pion^+)}\xspace}

\def\BptoRhopRho  {\decay{\Bp}{\rho^+\rho^0}}

\def\BdtoRhofznine  {\decay{\Bd}{\rho^0\ensuremath{f_0(980)}}}

\def\BtoA1Pi {\decay{\Bz}{a_1^+(\to \rhoz \pion^+)\pion^-}}

\newcommand{\eq}[1]{Eq.(\ref{equation : #1})}

\newcommand{\tab}[1]{Table~\ref{tab : #1}}

\newcommand{\fig}[1]{Figure~\ref{fig : #1}}

\def\fL {\ensuremath{f_{\rm L}}\xspace}

\def\swave {S-wave\xspace}
\def\pwave {P-wave\xspace}
\def\dwave {D-wave\xspace}

\begin{document}

\renewcommand{\thefootnote}{\fnsymbol{footnote}}
\setcounter{footnote}{1}


\begin{titlepage}
\pagenumbering{roman}

\vspace*{-1.5cm}
\centerline{\large EUROPEAN ORGANIZATION FOR NUCLEAR RESEARCH (CERN)}
\vspace*{1.5cm}
\hspace*{-0.5cm}
\begin{tabular*}{\linewidth}{lc@{\extracolsep{\fill}}r}
\ifthenelse{\boolean{pdflatex}}
{\vspace*{-2.7cm}\mbox{\!\!\!\includegraphics[width=.14\textwidth]{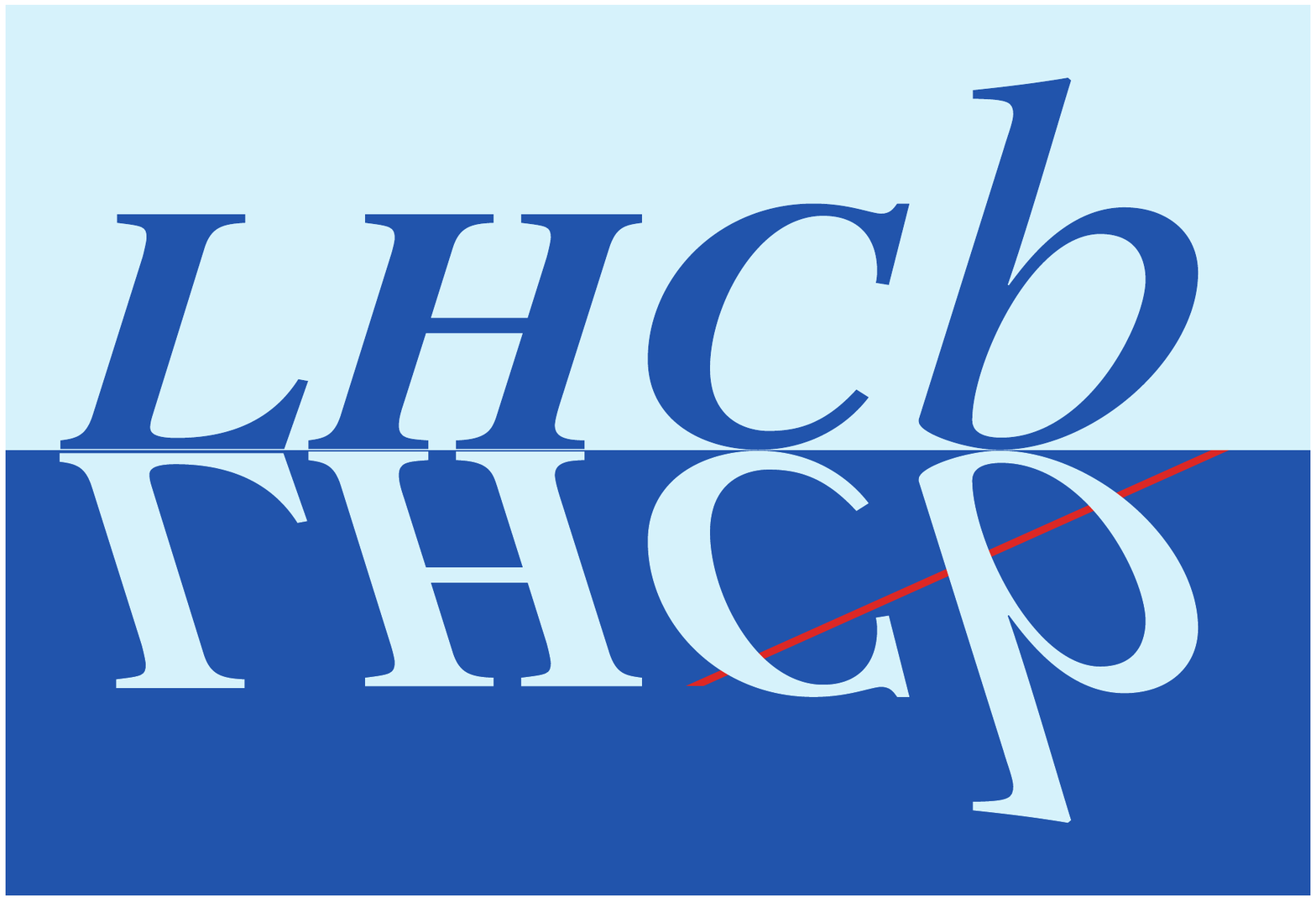}} & &}%
{\vspace*{-1.2cm}\mbox{\!\!\!\includegraphics[width=.12\textwidth]{lhcb-logo.eps}} & &}%
\\
 & & CERN-PH-EP-2015-077 \\  
 & & LHCb-PAPER-2015-006 \\  
 & & \today \\ 
\end{tabular*}

\vspace*{4.0cm}

{\bf\boldmath\huge
\begin{center}
Observation of the ${B^0 \to \rho^0 \rho^0}$ decay from an amplitude analysis of ${B^0 \to (\pi^+\pi^-)(\pi^+\pi^-)}$ decays
\end{center}
}

\vspace*{1.2cm}

\begin{center}
The LHCb collaboration\footnote{Authors are listed at the end of this letter.}
\end{center}

\vspace{\fill}

\begin{abstract}

Proton-proton collision data recorded in 2011 and 2012 by the \lhcb experiment, 
co\-rres\-pon\-ding to an integrated luminosity of 
3.0\invfb, are a\-na\-lysed to search for the charmless 
${B^0 \to \rho^0 \rho^0}$ decay.  
More than 600 ${\B^0 \to (\pi^+\pi^-)(\pi^+\pi^-)}$ signal decays are selected and used to perform an amplitude analysis, 
under the assumption of no CP violation in the decay, 
from which
the ${B^0 \to \rho^0 \rho^0}$ decay is observed for the first time with 7.1 standard deviations significance.
The fraction of ${B^0 \to \rho^0 \rho^0}$ decays yielding a longitudinally polarised final state is measured to be $\fL = 0.745^{+0.048}_{-0.058} ({\rm stat}) \pm 0.034 ({\rm syst})$.
The ${B^0 \to \rho^0 \rho^0}$ branching fraction, using the ${\B^0 \to \phi K^*(892)^{0}}$ decay as reference, is also reported as ${\BF(B^0 \to \rho^0 \rho^0) = (0.94 \pm 0.17 ({\rm stat}) \pm 0.09 ({\rm syst}) \pm 0.06 ({\rm BF}))  \times 10^{-6}}$.

\end{abstract}


\vspace*{1.5cm}

\begin{center}
  Published in Phys.~Lett.~B
\end{center}

\vspace{\fill}

{\footnotesize 
\centerline{\copyright~CERN on behalf of the \lhcb collaboration, license \href{http://creativecommons.org/licenses/by/4.0/}{CC-BY-4.0}.}}
\vspace*{2mm}

\end{titlepage}


\newpage
\setcounter{page}{2}
\mbox{~}
%
%
%
%

\cleardoublepage


\renewcommand{\thefootnote}{\arabic{footnote}}
\setcounter{footnote}{0}



\pagestyle{plain} 
\setcounter{page}{1}
\pagenumbering{arabic}


%

\section{Introduction}
\label{sec : Introduction}

The study of \B meson decays to $\rho\rho$ final states
provides the most powerful constraint to date for the Cabibbo-Kobayashi-Maskawa (CKM) angle 
$\alpha \equiv {\rm arg} \left[ (V_{td}V^*_{tb})/(V_{ud}V^*_{ub}) \right]$~\cite{Gronau:1990ka, Charles:2015gya, Bevan:2013kaa}. 
Most of the physics information is provided by the  decay  $B^0 \to \rho^+\rho^-$ as measured at the $e^+ e^-$ colliders at the $\Upsilon ({\rm 4S})$ resonance~\cite{Aubert:2007nua,Abe:2007ez},\footnote{Charge conjugation is implicit throughout the text unless otherwise stated.} for which the dominant decay amplitude, involving the emission of a \textsl{W} boson only (tree), 
exhibits a phase difference
that can be interpreted as the sum of the CKM angles $\beta +\gamma = \pi - \alpha$ in the Standard Model. The subleading amplitude associated with the exchange of a \textsl{W} boson and a quark (penguin) must be determined in order to interpret the electroweak phase difference in terms of the angle $\alpha$. This is realised by means of an isospin analysis involving the companion modes  
\BptoRhopRho~\cite{Zhang:2003up,Aubert:2006sb} and \BdtoRhoRho~\cite{Aubert:2008au,Adachi:2012cz}.\footnote{$\rho^0$ stands for $\rho^0(770)$ throughout the text.} 
In particular, the smallness of the amplitude of the latter leads to a better constraint on $\alpha$.

The~\babar and~\belle experiments reported evidence for the \BdtoRhoRho decay~\cite{Aubert:2008au,Adachi:2012cz} 
with an average branching fraction of ${\BR(\BdtoRhoRho) = (0.97 \pm 0.24) \times 10^{-6}}$~\cite{Aubert:2008au,Adachi:2012cz}. Despite small observed signal yields, each experiment measured the fraction \fL of decays yielding a longitudinally polarised final state through an angular analysis. The \belle collaboration did not find evidence for polarisation, $\fL = 0.21^{+0.22}_{-0.26}$~\cite{Adachi:2012cz}, while the \babar experiment measured a mostly longitudinally polarised decay, $\fL = 0.75^{+0.12}_{-0.15}$~\cite{Aubert:2008au}. These results differ at the level of $2.0$ standard deviations. 
The large \lhcb data set may shed light on this discrepancy. In addition, \lhcb may
confirm the hint of \BdtoRhofznine decays reported by \belle~\cite{Adachi:2012cz}. 
Measurements of the \BdtoRhoRho branching fraction and longitudinal polarisation fraction at \lhcb can be used as 
inputs in the determination of $\alpha$~\cite{Charles:2015gya,Bevan:2013kaa}.

This work focuses on the search and study of the \BtoPiPiPiPi decay in which the two \PiPi pairs are selected in the low invariant mass range ($<1100$ \mevcc). 
The \BdtoRhoRho is expected to dominate the \PiPi mass spectrum. 
The \PiPi combinations can actually emerge from \swave non-resonant and resonant contributions or other P- or \dwave resonances interfering with the signal. Hence, the determination of the \BdtoRhoRho yields requires a two-body mass and angular analysis,
from which the fraction of the longitudinally polarised final state can be measured.

The branching fraction is measured relative to the \BdtoPhiKst mode. The \BdtoPhiKst decay, which results in four light mesons in the final state, is similar to the signal, 
thus allowing for a cancellation of the uncertainties in the ratio of  
selection efficiencies.

\section{Data sets and selection requirements}
\label{sec : DataSelection} 

The analysed data correspond to an integrated luminosity of 1.0\invfb~and 2.0\invfb~from $pp$ collisions recorded at a 
centre-of-mass energy of 7\tev, collected in 2011, and 8\tev, collected in 2012, by the \lhcb experiment at \cern.

The \lhcb detector~\cite{Alves:2008zz,LHCb-DP-2014-002} is a single-arm forward
spectrometer covering the \mbox{pseudorapidity} range $2<\eta <5$,
designed for the study of particles containing \bquark or \cquark
quarks. It includes a high-precision tracking system
consisting of a silicon-strip vertex detector surrounding the $pp$
interaction region~\cite{LHCb-DP-2014-001}, a large-area silicon-strip detector located
upstream of a dipole magnet with a bending power of about
$4{\rm\,Tm}$, and three stations of silicon-strip detectors and straw
drift tubes~\cite{LHCb-DP-2013-003} placed downstream of the magnet.
The tracking system provides a measurement of momentum, \ptot, of charged particles with
a relative uncertainty that varies from 0.5\% at low momentum to 1.0\% at 200\gevc.
The minimum distance of a track to a primary vertex, the impact parameter, is measured with a resolution of $(15+29/\pt)\mum$,
where \pt is the component of the momentum transverse to the beam, in \gevc.
Different types of charged hadrons are distinguished using information
from two ring-imaging Cherenkov (RICH) detectors~\cite{LHCb-DP-2012-003}.
Photons, electrons and hadrons are identified by a calorimeter system consisting of
scintillating-pad and preshower detectors, an electromagnetic
calorimeter and a hadronic calorimeter. Muons are identified by a
system composed of alternating layers of iron and multiwire
proportional chambers~\cite{LHCb-DP-2012-002}.
The online event selection is performed by a trigger~\cite{LHCb-DP-2012-004},
which consists of a hardware stage, based on information from the calorimeter and muon
systems, followed by a software stage, which applies a full event
reconstruction.

In this analysis two categories of events 
that pass the hardware trigger stage are considered: 
those where the trigger decision is satisfied  
by the signal \bquark-hadron decay products ({\rm TOS}) 
and those where only the other activity in the event determines the trigger decision ({\rm TIS}). 
The software trigger requires a two-, three- or four-track secondary vertex with large transverse momenta of charged particles and a significant displacement 
from the primary $pp$ interaction vertices~(PVs). At least one charged particle should have $\pt >1.7\gevc$ and 
is required to be inconsistent with originating from any primary interaction. 
A multivariate algorithm~\cite{BBDT} is used for the identification of secondary vertices consistent
 with the decay of a \bquark hadron.

Further selection criteria are applied offline to reduce the number of background events with respect to the signal. 
The \PiPi candidates must have transverse momentum larger than 600\mevc, with at least one charged decay product with $\pt > 1000 \mevc$. 
The two \PiPi pairs are then combined to form a \Bd candidate with a good vertex quality and transverse momentum larger than 2500\mevc.
The invariant mass of each pair of opposite-charge pions forming the \Bd candidate is required
to be in the range 300--1100\mevcc. 
The identification of the final-state particles (PID) is performed with dedicated neural-networks-based discriminating variables 
that combine information from the RICH detectors and other properties of the event~\cite{LHCb-DP-2012-003}.  
The combinatorial background is further suppressed with multivariate discriminators based on a boosted decision tree
algorithm~(BDT)~\cite{Breiman,AdaBoost}. The BDT is trained with simulated \BdtoRhoRho (where $\rhoz \to \pion^+ \pion^-$) events as signal sample 
and candidates reconstructed with four-body mass in excess of 5420\mevcc as background sample. 
The discriminating variables are based on the kinematics of the $B$ decay candidate ($B$ \pt and the minimum \pt of the two $\rho^0$ candidates) and on geometrical vertex measurements (quality of the $B$ candidate vertex, impact parameter significances of the daughters, $B$ flight distance significance and $B$ pointing to the primary vertex).
The optimal thresholds for the BDT and PID discriminating variables
are determined simultaneously by 
means of a frequentist estimator for which no hypothesis on the signal yield is assumed~\cite{Punzi:2003bu}. 
The \Bd meson 
candidates are accepted in the mass range 5050--5500\mevcc. 

The normalisation mode \BdtoPhiKst is selected with similar criteria,  requiring in addition that the invariant mass of
the \KPi candidate is found in a range 
of $\pm$150\mevcc around the known value of the \Kstarz meson mass~\cite{PDG2014} and the invariant mass of the \KK pair is in a range 
of $\pm$15\mevcc centred at the known value of the $\phi$ meson mass~\cite{PDG2014}. 
A sample enriched in $\Bd \to (\kaon^+\pion^-)(\pion^+\pion^-)$~events is selected using the same ranges in \PiPi and \KPi masses to estimate the background with one misidentified kaon.

The presence of \PiPi pairs originating from \jpsi, $\chi_{c0}$ and $\chi_{c2}$ charmonia decays is vetoed by requiring the
invariant masses $M$ of all possible \PiPi pairs to be $|M-M_{0}| > 30 \mevcc$, where $M_{0}$ stands for the corresponding 
known values of the \jpsi,
$\chi_{c0}$ and $\chi_{c2}$ meson masses~\cite{PDG2014}. Similarly, the decays $\Dz \to \kaon^- \pion^+$ and $\Dz \to \pion^+ \pion^-$ are vetoed by 
requiring the corresponding invariant masses to differ by 25\mevcc or more from the known \Dz meson mass~\cite{PDG2014}. To reduce contamination from other charm backgrounds and from the \BtoA1Pi decay, the invariant
mass of any three-body combination in the event is required to be larger than 2100\mevcc. 

Simulated \BdtoRhoRho and \BdtoPhiKst decays are also used for  
determining the relative reconstruction efficiencies.
The $pp$ collisions are generated using
\pythia~\cite{Sjostrand:2007gs} with a specific \lhcb~configuration~\cite{LHCb-PROC-2010-056}. Decays of hadronic particles 
are described by \evtgen~\cite{Lange:2001uf}.
The interaction of the generated particles with the detector and its response are implemented using the \geant toolkit~\cite{Allison:2006ve, *Agostinelli:2002hh} 
as described in Ref.~\cite{LHCb-PROC-2011-006}.

\section{Four-body mass fit}
\label{sec : four-body-fit}

The four-body mass spectrum $M (\pi^+\pi^-)(\pi^+\pi^-)$ is fit with  
an unbinned extended likelihood.  The fit is performed simultaneously for the two data taking periods 
together with the normalisation channel $M (K^+K^-)(K^+\pi^-)$ and PID misidentification control channel $M (K^+\pi^-)(\pi^+\pi^-)$ mass spectra. The four-body invariant mass models account for \Bd and possible \Bs signals,
combinatorial backgrounds, signal cross-feeds  
and background contributions arising from partially reconstructed $b$-hadron decays in which one or more particles  
are not reconstructed.

The \Bd and \Bs meson shapes are modelled with a modified Crystal Ball distribution~\cite{Skwarnicki:1986xj}. A second power-law tail is added on the high-mass side of the signal shape to account for imperfections of the tracking system. The model parameters are determined from a simultaneous fit of simulated signal events that fulfill the trigger, reconstruction and selection chain, for each data taking period. The values of the tail parameters are identical for the \Bd and \Bs mesons. Their mass difference is constrained to the value from Ref.~\cite{PDG2014}. The mean and width of the modified Crystal Ball function are free parameters of the fit to the data.

The combinatorial background in each four-body spectrum is described by an exponential function where the slope is allowed to vary in the fit. 

The misidentification of one or more final-state hadrons may result in a fully reconstructed background contribution 
to the corresponding signal spectrum, 
denoted signal cross-feed. The magnitude of  the branching fractions of the signal and control modes as well as the two-body mass selection criteria make these signal cross-feeds negligible, with one exception:   
the misidentification of the kaon of the decay $\Bd \to (K^+\pi^-)(\pi^+\pi^-)$ as a pion yields a significant contribution in the  $M (\pi^+\pi^-)(\pi^+\pi^-)$ mass spectrum. 
The mass shape of $\Bd \to (K^+\pi^-)(\pi^+\pi^-)$ decays reconstructed as $\Bd \to (\pi^+\pi^-)(\pi^+\pi^-)$ is modelled by a Crystal Ball function, whose parameters are determined from simulated events. The  yield of this signal cross-feed is allowed to vary in the fit. The measurement of the actual number of reconstructed $\Bd \to (K^+\pi^-)(\pi^+\pi^-)$ events multiplied by the data-driven estimate of the misidentification efficiency 
is consistent with the measured yield.

The partially reconstructed background is modelled by an ARGUS function~\cite{Albrecht:1990cs} convolved with a Gaussian function accounting for resolution effects. Various mass shape parameterisations are examined. The best fit is obtained when the endpoint of the ARGUS function is fixed to the value expected when one pion is not attributed to the decay. The other shape parameters of the ARGUS function are free parameters of the fit, common to the two data taking periods. The floating width parameter of the signal mass shape is constrained to be equal to the width of the Gaussian function used in the convolution. 

~\fig{PaperFitResults} displays the $M (\pi^+\pi^-)(\pi^+\pi^-)$ and $M (K^+K^-)(K^+\pi^-)$ spectra with the fit results overlaid.  The signal event yields are shown in~\tab{PaperFitResults}. 
\begin{figure}[!t]
  \begin{center}
  \ifthenelse{\boolean{pdflatex}}{ 	   	  			        
    \includegraphics*[width=0.495\textwidth]{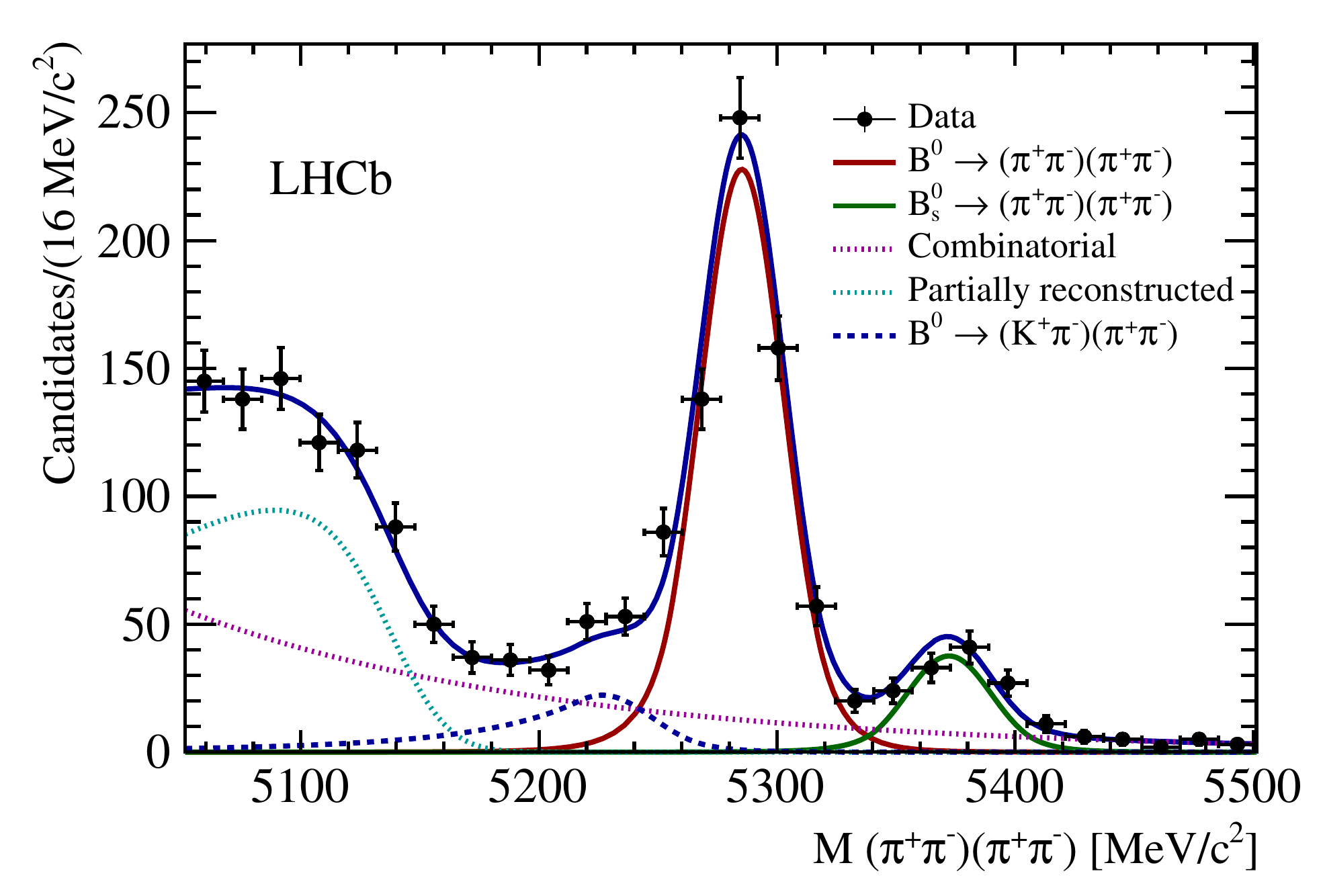}   
    \includegraphics*[width=0.495\textwidth]{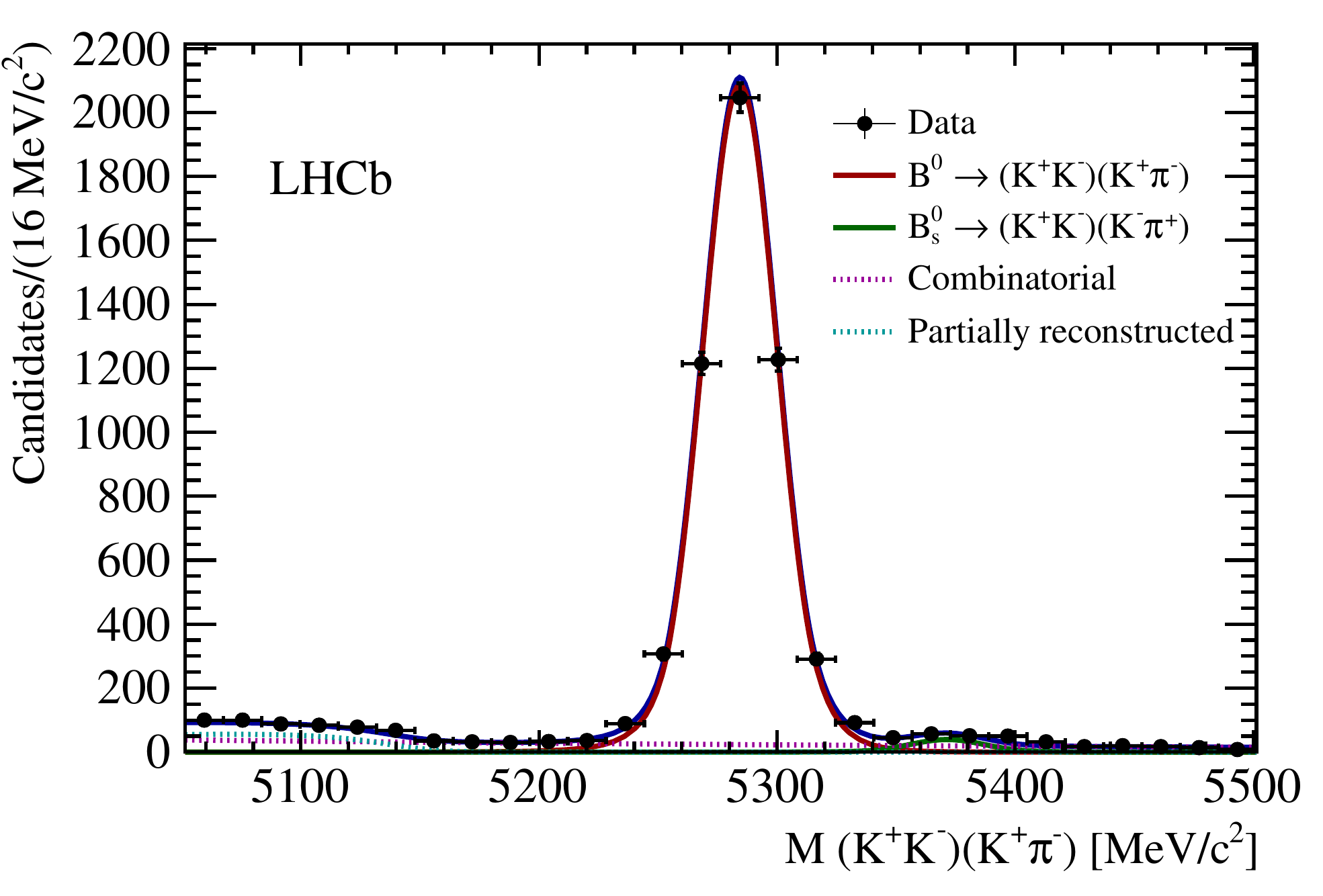}\\    			        
 }{
   }
  \end{center}
  \caption{\footnotesize Reconstructed invariant mass spectrum of (left)\PiPiPiPi~and (right)$(K^+K^-)(K^+\pi^-)$. The data are represented by the black dots. The fit is represented by the solid blue line, 
  the \Bd signal by the solid red line and the \Bs by the solid green line. 
  The combinatorial background is represented by the pink dotted line, the partially reconstructed background by the cyan dotted line and the cross-feed by the dark blue dashed line. 
 }
\label{fig : PaperFitResults}
\end{figure}
Aside from the prominent signal of the $\Bd \to (\pi^+\pi^-)(\pi^+\pi^-)$ decays, the decay mode $\Bs \to (\pi^+\pi^-)(\pi^+\pi^-)$ is observed with a statistical significance of more than 10 standard deviations. The statistical significance is evaluated by taking the ratio of the likelihood of the nominal fit and of the fit with the signal yield fixed to zero.

\begin{table}[!t]
\centering
\caption{Yields from the simultaneous fit for the 2011 and 2012 data sets. The first and second uncertainties are the statistical and systematic contributions, respectively.}
\begin{footnotesize}
\begin{tabular}{l  c   c}
\hline
\hline
Decay mode & Signal yields 2011 & Signal yields 2012 \\
\hline
$\BtoPiPiPiPi$                      &  $\phantom{0}185 \pm 15 \pm 4$ & $\phantom{0}449 \pm 24 \pm \phantom{0}7$ \\
$\BtoKPiPiPi$                      &  $1610 \pm 42 \pm 5$ & $3478 \pm 62 \pm 10$ \\
$\BtoKKKPi$                         &  $1513 \pm 40 \pm 8$ & $3602 \pm 62 \pm 10$ \\
\hline
$\BstoPiPiPiPi$                     & $\phantom{00}30 \pm \phantom{0}7 \pm 1$ & $\phantom{00}71 \pm 11 \pm \phantom{0}1$ \\
$\BstoKPiPiPi$                     & $\phantom{00}40 \pm 10 \pm 3$ & $\phantom{00}96 \pm 14 \pm \phantom{0}6$ \\
$\BstoKKKPi$                        & $\phantom{00}42 \pm 10 \pm 3$ & $\phantom{00}66 \pm 13 \pm \phantom{0}4$ \\
\hline
\hline
\end{tabular}
\label{tab : PaperFitResults}
\end{footnotesize}
\end{table}

A systematic uncertainty due to the fit model is associated to the measured yields. The dominant uncertainties arise from the knowledge of the signal and signal cross-feed shape parameters determined from simulated events. Several pseudoexperiments are generated while varying the shape parameters within their uncertainties, and the systematic uncertainties on the yields are estimated from the differences in results with respect to the nominal fit.

\section{Amplitude analysis}
\label{sec : angular-analysis}


An amplitude analysis is used to determine the vector-vector (VV) contribution \BdtoRhoRho
by using two-body mass spectra and angular variables.
The four-body mass spectrum is first analysed with the $\sPlot$ technique~\cite{Pivk:2004ty} to subtract statistically the background under the $\Bd \to (\pi^+\pi^-)(\pi^+\pi^-)$ signal.

For the two-body mass spectra, contributions from resonant and non-resonant scalar ($S$), resonant  vector ($V$) and tensor ($T$) components are considered in the amplitude fit model through complex mass propagators, $M(m_i)$,
where the label $i=1,2$ are the first and second pion pairs, which are assigned randomly in every decay since they are indistinguishable.
The \pwave lineshape model comprises the $\rho^0$ meson, described using the Gounaris-Sakurai parameterisation $M_{\rho}(m_i)$~\cite{Gounaris:1968mw},
and the $\omega$ meson, parameterised with a relativistic spin-1 Breit-Wigner $M_{\omega}(m_i)$. 
The \dwave lineshape  $M_{f_2}(m_i)$ accounts  
for the $f_2(1270)$, modelled with a relativistic spin-2 Breit-Wigner.
The \swave model includes the  
$f_0(980)$ propagator $M_{f(980)}(m_i)$, described using a Flatt\'e parameterisation~\cite{Flatte:1976xv,Flatte:1976xu}, 
and a low-mass component. The latter includes the broad low-mass resonance $f_0(500)$ and a non-resonant contributions, which are jointly modelled in the framework of the $K$--matrix formalism~\cite{ANDP:ANDP19955070504} and referred as $M_{(\pi\pi)_0}(m_i)$.
Following the $K$--matrix formalism, the amplitude for the low-mass $\pi^+\pi^-$ \swave can be written  as 
\begin{eqnarray}
A(m) \propto \frac{\hat{K}}{1-i\rho \hat{K}} ,
\end{eqnarray}
with
\begin{eqnarray}
&\hat{K}& \equiv \hat{K}_{{\rm res}} + \hat{K}_{{\rm non-res}} =
\frac{m_0\Gamma(m)}{(m_0^2-m^2)\rho(m)}+\kappa , \\
&\rho(m)& = 2 \left( \frac{q(m)}{m} \right) ,
\end{eqnarray}
where $\kappa$ is measured to be $-0.07 \pm 0.24$ from a fit to the inclusive $\pi^+\pi^-$ mass distribution and  
$m_0$ and $\Gamma$ are the nominal mass and mass-dependent width of 
the $f_0(500)$, 
as determined in Ref.~\cite{Muramatsu:2002jp}.  
The functions $\rho(m)$ and $q(m)$, defined in Ref.~\cite{ANDP:ANDP19955070504}, are the phase space factor and the relative momentum of a pion
in the $\rho^0$ centre-of-mass system.  
By convention, the phase of the $M_{(\pi\pi)_0}(m_i)$ mass propagator is set to zero at the $\rho^0$ nominal mass. 
\par

The signal sample is described by considering the dominant amplitudes of the signal decay.
The $B\to VV$ component contains the \BtoRhoRho and $B^0 \to \rho^0 \omega$ amplitudes. 
The $B\to VS$ component accounts for $B^0 \to \rho^0 \PiPi_0$ and $B^0
\to \rho^0 f_0(980)$ amplitudes and the $B\to VT$ contribution is limited to the purely longitudinal amplitude of the $B^0 \to \rho^0 f_2(1270)$ transition. Because of the broad natural width of the $a_1^{\pm}$ particle, a small contamination from the decays $B^0 \to a_1^{\pm}\pi^{\mp}$ remains in the sample. 
This contribution
with $a_1^{\pm} \to \rho^0 \pi^{\pm}$ in S-wave is considered along with its interference with the other amplitudes. 
This is done by introducing the \CP-even eigenstate from the linear combination of individual amplitudes of the decays
$B^0 \to a_1^+ \pi^-$ and $B^0 \to a_1^- \pi^+$, as defined in Ref.~\cite{Bhattacharya:2013sga}.
The contribution of the decays $B^0\to \omega \omega$, $B^0\to f_0(980)f_0(980)$, $B^0 \to \omega S$, $B^0\to \omega T$, $B^0 \to f_2(1270) S$, $B^0\to f_2(1270) f_2(1270)$ and $B^0 \to (\rho^0 f_2(1270))_{\parallel,\perp}$ are assumed to be negligible,
where the $\parallel$ and $\perp$ subindices indicate the parallel and perpendicular amplitudes of the decay.
The choice of the baseline model was made prior to the measurement of the physical parameters of interest after comparing a set of alternative parameterisations according to a dissimilarity statistical test~\cite{Williams:2010vh}.

\begin{figure}[!t]
 \begin{center}
   \includegraphics*[width=8.cm]{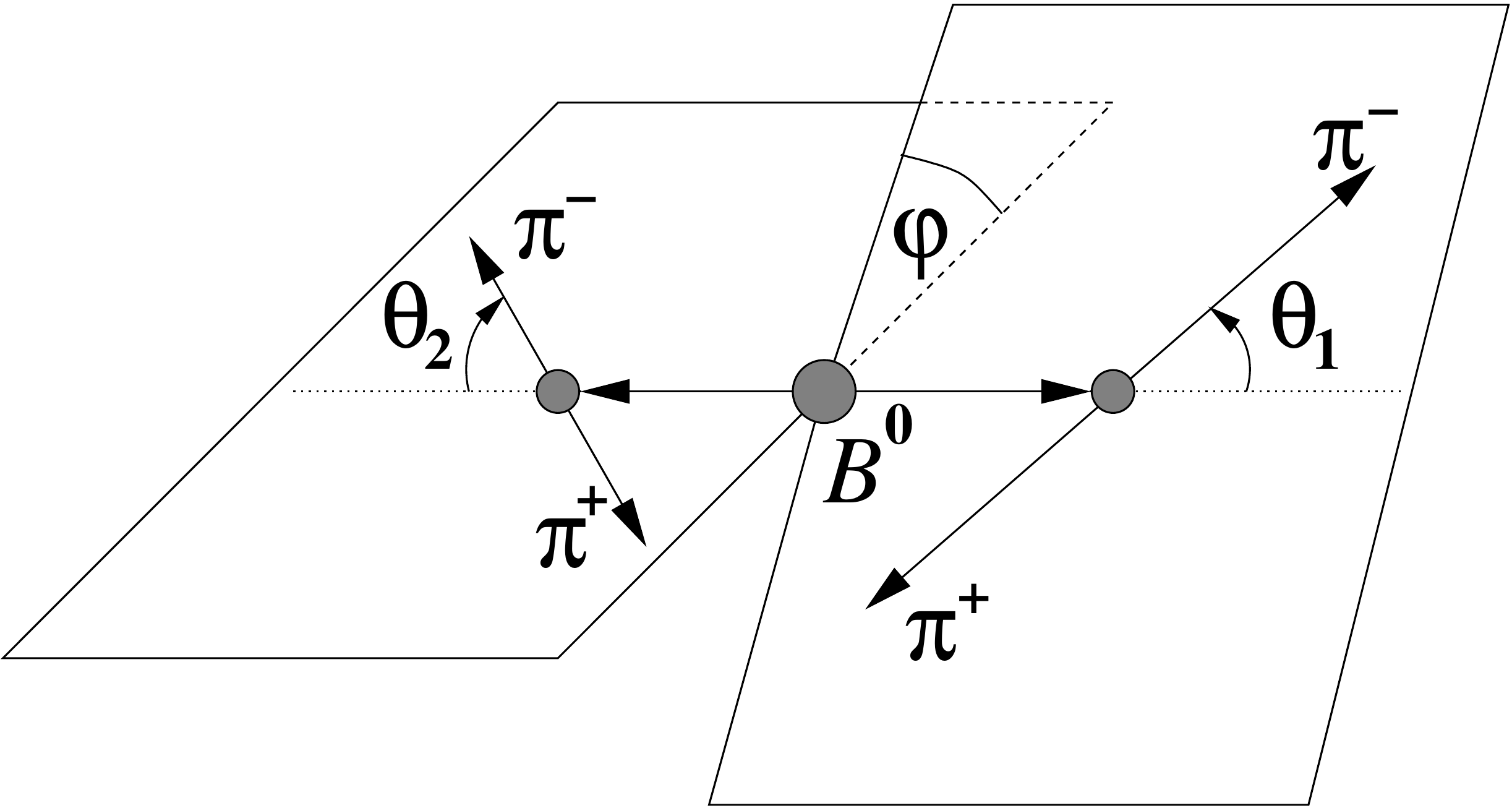}
 \end{center}
\caption{Helicity angles for the \PiPiPiPi system.}

 \label{fig : angles_RhoRho}
\end{figure}

The differential decay rate for \BtoPiPiPiPi decays at the $B^0$ production time $t=0$ is given by
\begin{equation}
\frac{{\rm d}^5\Gamma}{{\rm d}\cos\theta_1 \, {\rm d} \cos\theta_2 \, {\rm d} \varphi \, {\rm d} m_1^2 \, {\rm d} m_2^2}
\propto {\Phi}_4(m_1,m_2) \left|\sum_{i=1}^{11} A_i f_i(m_1,m_2,\theta_1,\theta_2,\varphi)\right|^2, 
\end{equation}
\noindent  where the variables $\theta_1$,  $\theta_2$ and $\varphi$ are the helicity angles, described in Fig.~\ref{fig : angles_RhoRho}, and ${\Phi}_4$ is the four-body phase space factor. The notations of  the complex amplitudes, $A_i$, and the expressions of their related angular distributions, $f_i$, are displayed in~\tab{ampterms}. 
The mass propagators included in the $f_i$ functions are normalised to unity in the fit range. 

\begin{table}[t!]
\caption{Amplitudes, $A_i$, \CP eigenvalues, $\eta_i$, and mass-angle distributions, $f_i$, of the \BtoPiPiPiPi model.
The indices ${ijkl}$ indicate the eight possible combinations of pairs of opposite-charge pions.
The angles $\alpha_{kl}$, $\beta_{ij}$ and $\Phi_{kl}$ are defined in Ref.~\cite{Aubert:2009ab}.
}
\begin{center}
\resizebox{\textwidth}{!}{%
\begin{tabular}{c c c} \hline\hline
$A_i$                         & $\eta_i$   & $f_i$ \\ \hline
$A_{\rho\rho}^0$              & $\phantom{-}1$ & $M_\rho(m_1)M_\rho(m_2)\cos\theta_1\cos\theta_2$ \\ 
$A_{\rho\rho}^{\parallel}$    & $\phantom{-}1$ & $M_\rho(m_1)M_\rho(m_2)\frac{1}{\sqrt{2}}\sin\theta_1\sin\theta_2\cos\varphi$ \\ 
$A_{\rho\rho}^{\perp}$        & $-1$ & $M_\rho(m_1)M_\rho(m_2)\frac{i}{\sqrt{2}}\sin\theta_1\sin\theta_2\sin\varphi$ \\ 
$A_{\rho\omega}^0$            & $\phantom{-}1$ & $\frac{1}{\sqrt{2}}[M_\rho(m_1)M_\omega(m_2)+M_\omega(m_1)M_\rho(m_2)]\cos\theta_1\cos\theta_2$ \\ 
$A_{\rho\omega}^{\parallel}$  & $\phantom{-}1$ & $\frac{1}{\sqrt{2}}[M_\rho(m_1)M_\omega(m_2)+M_\omega(m_1)M_\rho(m_2)]\frac{1}{\sqrt{2}}\sin\theta_1\sin\theta_2\cos\varphi$ \\ 
$A_{\rho\omega}^{\perp}$      & $-1$ & $\frac{1}{\sqrt{2}}[M_\rho(m_1)M_\omega(m_2)+M_\omega(m_1)M_\rho(m_2)]\frac{i}{\sqrt{2}}\sin\theta_1\sin\theta_2\sin\varphi$ \\ 
$A_{\rho(\pi\pi)_0}$          & $-1$ & $\frac{1}{\sqrt{6}}[M_\rho(m_1)M_{(\pi\pi)_0}(m_2)\cos\theta_1+M_{(\pi\pi)_0}(m_1)M_\rho(m_2)\cos\theta_2]$ \\ 
$A_{\rho f(980)}$             & $-1$ & $\frac{1}{\sqrt{6}}[M_\rho(m_1)M_{f(980)}(m_2)\cos\theta_1+M_{f(980)}(m_1)M_\rho(m_2)\cos\theta_2]$ \\ 
$A_{(\pi\pi)_0(\pi\pi)_0}$    & $\phantom{-}1$ & $M_{(\pi\pi)_0}(m_1)M_{(\pi\pi)_0}(m_2)\frac{1}{3}$ \\ 
$A_{\rho f_2}^0$              & $-1$ & $\sqrt{\frac{5}{24}} \left[M_{\rho}(m_1)M_{f_2}(m_2)\cos\theta_1(3\cos^2\theta_2-1)
 + M_{f_2}(m_1)M_{\rho}(m_2)\cos\theta_2(3\cos^2\theta_1-1)\right]$ \\
$A_{a_1\pi}^{S^+}$              & $\phantom{-}1$  & $\frac{1}{\sqrt{8}}\sum_{\{ijkl\}}\frac{1}{\sqrt{3}}M_{a_1}(m_{ijk})M_{\rho}(m_{ij})\left[ \cos\alpha_{kl}\cos\beta_{ik} + \sin\alpha_{kl}\sin\beta_{ik}\cos\Phi_{kl} \right]$ \\
\hline\hline
\end{tabular}
}
\label{tab : ampterms}
\end{center}
\end{table}

For the \CP
conjugated mode, $\Bdb \to \PiPi \PiPi$, the decay rate is obtained under the transformation $A_i \rightarrow \eta_i \overline{A}_i$, where $\eta_i$ is the \CP eigenvalue of the
\CP eigenstate $i$, shown in~\tab{ampterms}.

The untagged time-integrated decay rate of \Bd and \Bdb to four pions, assuming no \CP violation, can be written as
\begin{eqnarray}
\label{eq : decayrate2}
\frac{{\rm d}^5(\Gamma+\overline{\Gamma})}{{\rm d}\cos\theta_1 \, {\rm d}\cos\theta_2 \, {\rm d}\varphi \, {\rm d} m_1^2
\, {\rm d} m_2^2} \propto \sum_{j=1}^{11} \sum_{i\leq j} \Real[A_iA_j^* f_if_j^{*}](2-\delta_{ij})(1+\eta_i\eta_j) {\Phi}_4(m_1,m_2)  \; , 
\end{eqnarray}
\noindent where $\delta_{ij}=1$ when $i=j$ and $\delta_{ij}=0$ otherwise.


The efficiency of the selection of the final state $\Bd \to (\pi^+\pi^-)(\pi^+\pi^-)$ varies as a function of the helicity angles and the two-body invariant masses. 
To take into account variations in the efficiencies, four event categories $k$ are defined according to their hardware trigger decisions (TIS or TOS) and data taking period (2011 and 2012).

The acceptance is accounted for through the 
complex integrals 
\begin{equation}
\label{equation : n_weights}
\omega^k_{ij} =
\int\epsilon(\theta_1,\theta_2,\varphi,m_1,m_2)f_if_j^{*}(2-\delta_{ij}){\Phi}_4(m_1,m_2)\deriv\cos\theta_1\,\deriv\cos\theta_2\,\deriv\varphi
\,\deriv m_1^2\,\deriv m_2^2, 
\end{equation}
\noindent where $f_i$ are the functions
given in \tab{ampterms} and $\epsilon$ the overall efficiency. The integrals are computed with simulated events of each of the
four considered categories, selected with the same
criteria as those applied to data,
following the method described in Ref.~\cite{TristansThesis}.
The coefficients $\omega^k_{ij}$ are used to
determine the efficiency  
and to build
a probability density function for each category, which is defined as
\begin{equation}
S^k (m_1,m_2,\theta_1,\theta_2,\varphi) = \frac{\sum_{j=1}^{11}\sum_{i\leq j} \Real[
A_iA_j^* f_i
f_j^{*}](2-\delta_{ij})(1+\eta_i\eta_j) {\Phi}_4(m_1,m_2)}{\sum_{j=1}^{11}\sum_{i\leq j} \Real[
A_iA_j^* \omega^k_{ij}](1+\eta_i\eta_j) }. 
\end{equation}

The four event categories are used in the simultaneous unbinned maximum likelihood fit which depends on the $19$ free parameters indicated in~\tab{Paperfitdatanocpv}.

\begin{table}[!t]
\caption{Results of the unbinned maximum likelihood fit to the angular and two-body invariant mass distributions. The first uncertainty is statistical, the second systematic.}
\begin{center}
\resizebox{\textwidth}{!}{%
\begin{tabular}{c c c}
\hline\hline
Parameter & Definition & Fit result \\
\hline\hline
 $\fL$ & $|A_{\rho\rho}^0|^2/(|A_{\rho\rho}^0|^2+|A_{\rho\rho}^{\parallel}|^2+|A_{\rho\rho}^{\perp}|^2)$ & $\phantom{-}0.745^{+0.048}_{-0.058}\pm 0.034$ \\
 $f_{\parallel}^{\prime}$ & $|A_{\rho\rho}^{\parallel}|^2/(|A_{\rho\rho}^{\parallel}|^2+|A_{\rho\rho}^{\perp}|^2)$ & 
$\phantom{-}0.50 \pm 0.09 \pm 0.05\phantom{0}$ \\
 $\delta_{\parallel}-\delta_0$ & $\arg(A_{\rho\rho}^{\parallel}A_{\rho\rho}^{0*})$ & 
$\phantom{-}1.84 \pm 0.20 \pm 0.14\phantom{0}$ \\
\hline
 $F_{\rho (\pi\pi)_0}$ & $|A_{\rho (\pi\pi)_0}|^2/(|A_{\rho\rho}^0|^2+|A_{\rho\rho}^{\parallel}|^2+|A_{\rho\rho}^{\perp}|^2)$ & 
$\phantom{-}0.30\phantom{0}^{+0.11\phantom{0}}_{-0.09\phantom{0}} \pm 0.08\phantom{0}$ \\
 $F_{\rho f(980)}$     & $|A_{\rho f(980)}|^2/(|A_{\rho\rho}^0|^2+|A_{\rho\rho}^{\parallel}|^2+|A_{\rho\rho}^{\perp}|^2)$ &    $\phantom{-}0.29\phantom{0}^{+0.12\phantom{0}}_{-0.09\phantom{0}} \pm 0.08\phantom{0}$ \\
 $F_{(\pi\pi)_0(\pi\pi)_0}$ & $|A_{(\pi\pi)_0(\pi\pi)_0}|^2/(|A_{\rho\rho}^0|^2+|A_{\rho\rho}^{\parallel}|^2+|A_{\rho\rho}^{\perp}|^2)$ &    $\phantom{-}0.21\phantom{0}^{+0.06\phantom{0}}_{-0.04\phantom{0}} \pm 0.08\phantom{0}$ \\
 $\delta_{\perp}-\delta_{\rho (\pi\pi)_0}$ & $\arg(A_{\rho\rho}^{\perp}A^*_{\rho(\pi\pi)_0})$ & $-1.13\phantom{0}^{+0.33\phantom{0}}_{-0.22\phantom{0}} \pm 0.24\phantom{0}$ \\
 $\delta_{\perp}-\delta_{\rho f(980)}$ &$\arg(A_{\rho\rho}^{\perp}A^*_{\rho f(980)})$ & 
$\phantom{-}1.92 \pm 0.24 \pm 0.16\phantom{0}$ \\
 $\delta_{(\pi\pi)_0(\pi\pi)_0}-\delta_0$ & $\arg(A_{(\pi\pi)_0(\pi\pi)_0}A_{\rho\rho}^{0*})$ & $\phantom{-}3.14\phantom{0}^{+0.36\phantom{0}}_{-0.38\phantom{0}} \pm 0.39\phantom{0}$ \\
\hline
 $F_{\rho \omega}$ & 
 $(|A_{\rho\omega}^{0}|^2+|A_{\rho\omega}^{\parallel}|^2+|A_{\rho\omega}^{\perp}|^2)/(|A_{\rho\rho}^0|^2+|A_{\rho\rho}^{\parallel}|^2+|A_{\rho\rho}^{\perp}|^2)$ & 
$\phantom{-}0.025^{+0.048}_{-0.022} \pm 0.020$ \\
 $f^{\rho \omega}_L$ & $|A_{\rho\omega}^0|^2/(|A_{\rho\omega}^0|^2+|A_{\rho\omega}^{\parallel}|^2+|A_{\rho\omega}^{\perp}|^2)$ & $\phantom{-}0.70\phantom{0}^{+0.23\phantom{0}}_{-0.60\phantom{0}} \pm 0.13\phantom{0}$ \\
 $f^{\rho \omega \prime}_{\parallel}$ & $|A_{\rho\omega}^{\parallel}|^2/(|A_{\rho\omega}^{\parallel}|^2+|A_{\rho\omega}^{\perp}|^2)$& $\phantom{-}0.97\phantom{0}^{+0.69\phantom{0}}_{-0.56\phantom{0}} \pm 0.15\phantom{0}$ \\
 $\delta^{\omega}_{0}-\delta_{0}$ & $\arg(A_{\rho\omega}^0A_{\rho\rho}^{0*})$ & 
$-2.56\phantom{0}^{+0.76\phantom{0}}_{-0.92\phantom{0}} \pm 0.22\phantom{0}$ \\
 $\delta^{\omega}_{\parallel}-\delta_0$ & $\arg(A_{\rho\omega}^{\parallel}A_{\rho\rho}^{0*})$ & $-0.71\phantom{0}^{+0.71\phantom{0}}_{-0.67\phantom{0}} \pm 0.32\phantom{0}$ \\
 $\delta^{\omega}_{\perp}-\delta_{\rho (\pi\pi)_0}$ & $\arg(A_{\rho\omega}^{\perp}A_{\rho (\pi\pi)_0}^*)$  & 
$-1.72 \pm 2.62 \pm 0.80\phantom{0}$ \\
\hline
 $F_{\rho f_2}^0$  & $|A_{\rho f_2}^0|^2/(|A_{\rho\rho}^0|^2+|A_{\rho\rho}^{\parallel}|^2+|A_{\rho\rho}^{\perp}|^2)$ &    $\phantom{-}0.01\phantom{0}^{+0.04\phantom{0}}_{-0.02\phantom{0}} \pm 0.03\phantom{0}$ \\
 $\delta_{\rho f_2}^0-\delta_{\rho (\pi\pi)_0}$ & $\arg(A_{\rho f_2}^0A_{\rho(\pi\pi)_0}^*)$ & 
$-0.56 \pm 1.48 \pm 0.80\phantom{0}$ \\
\hline

 $F_{a_1\pi}^{S^+}$  & $|A_{a_1\pi}^{S^+}|^2/(|A_{\rho\rho}^0|^2+|A_{\rho\rho}^{\parallel}|^2+|A_{\rho\rho}^{\perp}|^2)$ &  
$1.4\phantom{00}^{+1.0\phantom{00}}_{-0.7\phantom{00}} \phantom{0}^{+1.2\phantom{00}}_{-0.8\phantom{00}}$ \\
 $\delta_{a_1\pi}^{S^+}-\delta_{\rho (\pi\pi)_0}$ & $\arg(A_{a_1\pi}^{S^+} A_{\rho(\pi\pi)_0}^*)$ & 
$-0.09\phantom{0}^{+0.30\phantom{0}}_{-0.36\phantom{0}} \pm 0.38\phantom{0}$ \\
\hline\hline
 \end{tabular} 
}
\label{tab : Paperfitdatanocpv}
\end{center}
\end{table}


Systematic effects
are estimated by fitting with the 
angular model an ensemble of 1000 pseudoexperiments generated with 
the same number of events as observed in data. 
The biases are for the parameters of interest consistent with zero. 
A systematic uncertainty is assigned by taking 50\% of the fit bias or the uncertainty on the {\rm rms} when the latter is bigger
in order to account for possible statistical fluctuations.

Several model related uncertainties are envisaged.
The $B^0 \to a_1^{\pm}\pi^{\mp}$ angular model requires knowledge of the lineshape of the $a_1^{\pm}$ meson. The $a_1^{\pm}$ natural width is chosen to be 400\mevcc. 
The difference to the 
fit results obtained by varying the width from 250 to 600\mevcc is taken as the corresponding systematic uncertainty. 
In addition, a systematic uncertainty is obtained by introducing the \CP-odd component 
in the fit model of the decay amplitude $B^0 \to a_1^{\pm} \pi^{\mp}$ 
by fixing the relative amplitudes of $B^0 \to a_1^+ \pi^-$ and $B^0 \to a_1^- \pi^+$ components to the values measured in Ref.~\cite{Dalseno:2012hp}. 
Another source of uncertainty originates in the modelling of the low mass \PiPi \swave lineshape. The $f_0(500)$ mass and natural width uncertainties from Ref.~\cite{Muramatsu:2002jp} and the uncertainty on the parameter that quantifies the non-resonant contribution are propagated to the angular analysis parameters by generating and fitting 1000 pseudoexperiments in which these input values are varied according to a Gaussian distribution having their uncertainties as widths. 
The root mean square of the distribution of the results is assigned as a systematic uncertainty. 
The same strategy is followed to estimate the systematic uncertainties originating from the $\rho^0$, $f_0(500)$ and $\omega$ 
lineshape parameters. 

The uncertainty related to the background subtraction method is estimated by varying within their uncertainties the fixed parameters of the mass fit model and studying the resulting angular distributions and two-body mass spectra. The difference to the  
fit results is taken as a systematic uncertainty. An alternative subtraction of the background estimated from the high-mass sideband is performed, yielding compatible results.       

The knowledge of the acceptance model described in~\eq{n_weights} 
comes from a finite sample of simulated events. 
An ensemble of pseudoexperiments is generated by varying  
the acceptance weights according to their covariance matrix. The root mean square of the distribution of the results is assigned as a systematic uncertainty.

The resolution on the helicity angles is evaluated with pseudoexperiments resulting in a negligible systematic uncertainty. 
The systematic uncertainty related to the \PiPi mass resolution is estimated with
pseudoexperiments by introducing a smearing of the \PiPi mass.
Differences in the parameters between the fit with and without smearing are taken as a systematic uncertainty. 

~\tab{SummarySyst_ang} details the contributions to the systematic uncertainty in the measurement of the fraction of \BdtoRhoRho signal decays in the \BtoPiPiPiPi and its longitudinal polarisation fraction.  

\begin{table}[!t]
\caption{Relative systematic uncertainties on the longitudinal polarisation parameter, \fL, and the fraction of \BdtoRhoRho decays in the \BtoPiPiPiPi sample. The model uncertainty includes the three uncertainties below.}
\begin{center}
\begin{tabular}{l c c}
\hline \hline
Systematic effect           & Uncertainty on $\fL$ (\%)  & Uncertainty on $P(\BdtoRhoRho)$ (\%) \\
\hline \hline
Fit bias                    & $\phantom{<}$0.1       & $0.8$ \\
Model                       & $\phantom{<}$3.6       & $6.2$ \\
\hline
\small \quad $B^0\to a_1(1260)^+\pi^-$   & \small $\phantom{<}$1.2       & \small $1.1$ \\
\small \quad \swave lineshape            & \small $\phantom{<}$3.4       & \small $6.1$ \\
\small \quad Lineshapes                  & \small $<$0.1                 & \small $0.1$ \\
\hline
Background subtraction      & $\phantom{<}$0.1       & $0.5$ \\
Acceptance integrals        & $\phantom{<}$2.7       & $4.5$ \\
Angular/Mass resolution     & $\phantom{<}$0.8       & $1.5$ \\
\hline \hline
\end{tabular}
\end{center}
\label{tab : SummarySyst_ang}
\end{table}


The final results of the combined two-body mass and angular analysis 
are shown in~\fig{Paperfitdatanocpv} and~\tab{Paperfitdatanocpv}.
\begin{figure}[!t]
 \begin{center}
   \includegraphics*[width=5.2cm]{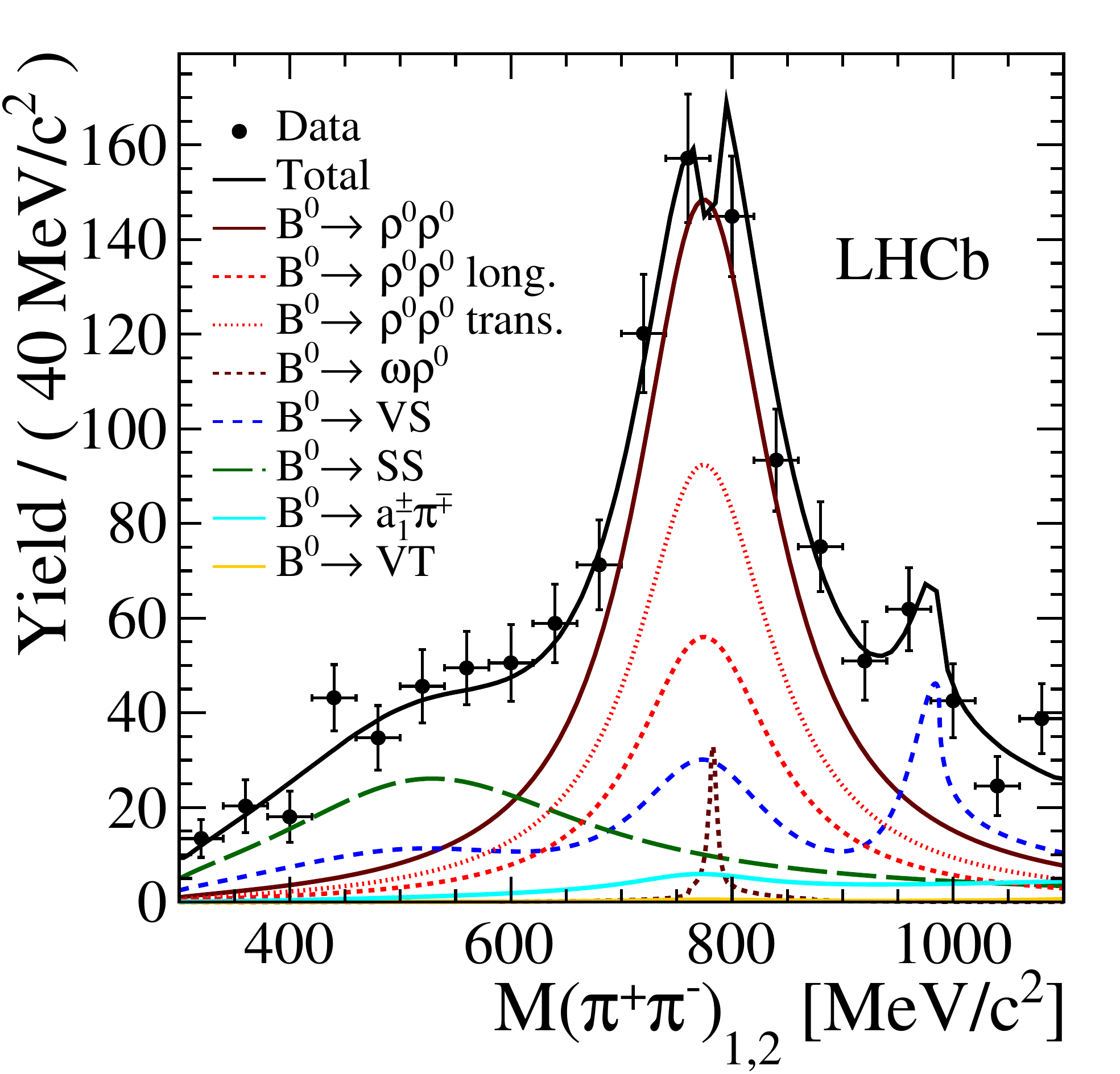}
   \includegraphics*[width=5.2cm]{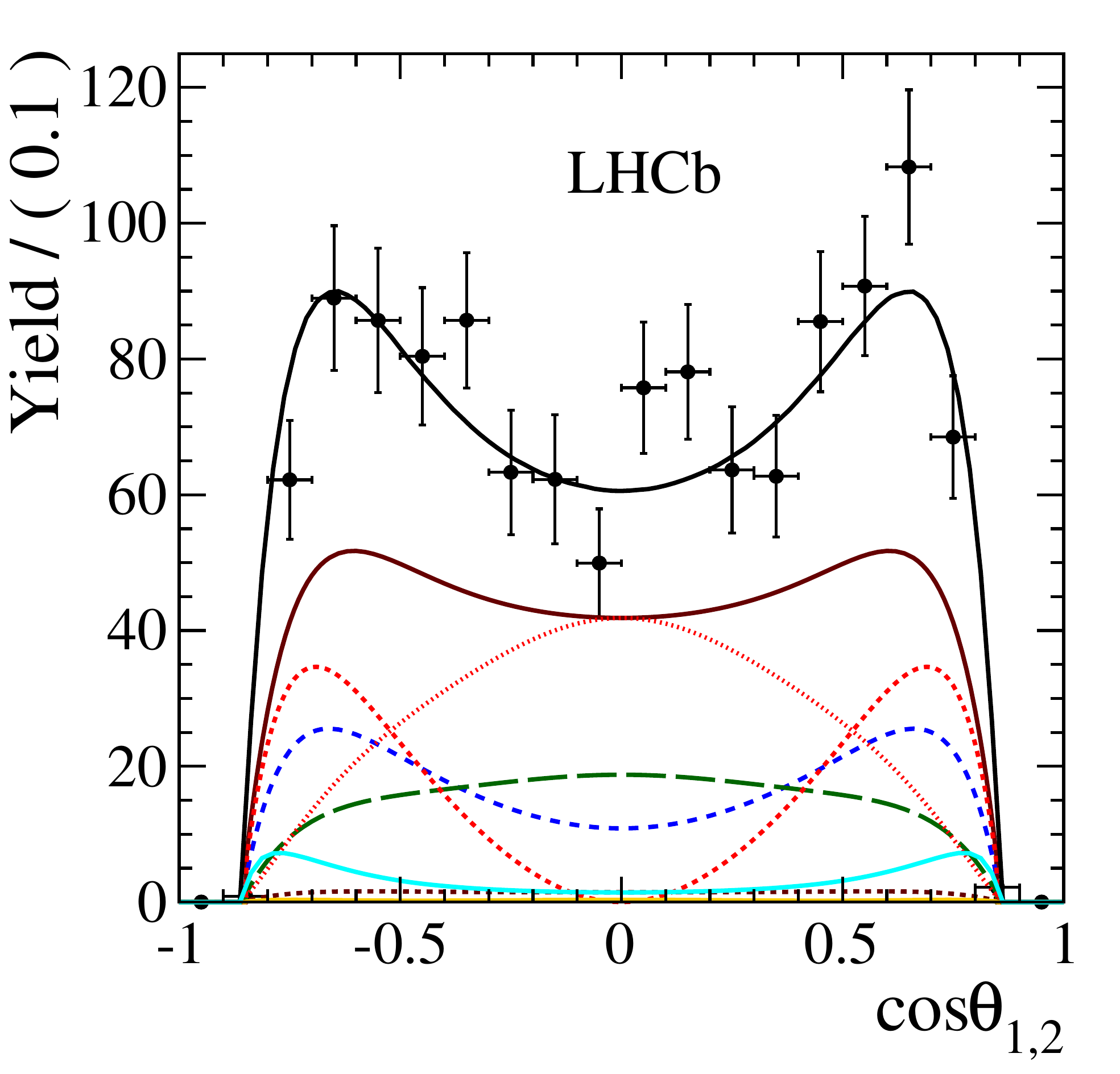}
   \includegraphics*[width=5.2cm]{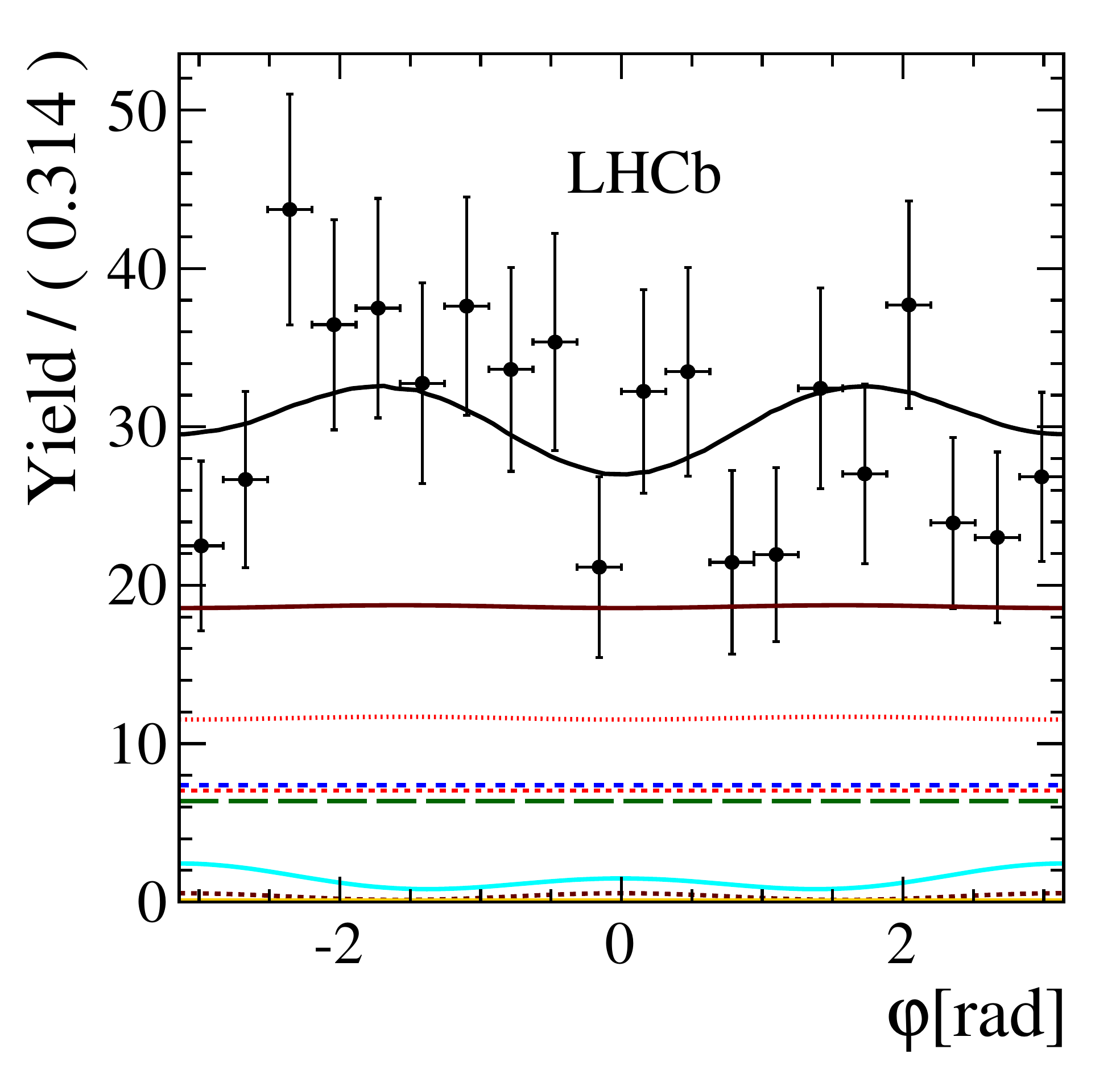}
 \end{center}
\caption{Background-subtracted $M\PiPi_{1,2}$, $\cos\theta_{1,2}$ and $\varphi$ 
distributions. The black dots correspond to the four-body background-subtracted
data and the black line is the projection of the fit model. The
specific decays $B^0\rightarrow \rho^0\rho^0$ (brown), $B^0 \to \omega \rho^0$
(dashed brown), $B^0\rightarrow VS$ (dashed blue), $B^0 \rightarrow SS$ (long
dashed green), $B^0\rightarrow VT$ (orange) and
$B^0 \to a_1^{\pm}\pi^{\mp}$ (light blue) are also displayed. The
$B^0\rightarrow \rho^0\rho^0$ contribution is split into longitudinal
(dashed red) and transverse (dotted red) components.
Interference contributions are only plotted for the total (black) model.
The efficiency for longitudinally polarized $B^0\rightarrow \rho^0\rho^0$ events is $\sim$5 times smaller than for the transverse component. 
}
\label{fig : Paperfitdatanocpv}
\end{figure}
The fit also allows for the extraction of the fraction of \BdtoRhoRho decays in the \BtoPiPiPiPi sample, defined as 
\begin{equation}
P(B^0\rightarrow \rho^0 \rho^0) =
\frac{\sum_{j=1}^{3}\sum_{i\leq j} \Real[
A_iA_j^* \omega_{ij} ]}{\sum_{j=1}^{11}\sum_{i\leq j} \Real[
A_iA_j^* \omega_{ij} ]}, 
\label{equation : purity_rho}
\end{equation}
\noindent which is 
\[
 {P(\BdtoRhoRho) = 0.619 \pm 0.072 \stat \pm 0.049 \syst}. 
\]
The \BdtoRhoRho signal significance is measured to be $7.1$ standard deviations. The significance is obtained by dividing the value of the purity by the quadrature of the statistical and systematic uncertainties.
No evidence for the $\Bd \to \rho^0 f_0(980)$ decay mode is obtained. 
The fraction of longitudinal polarisation of the $\BdtoRhoRho$ decay is measured to be 
\[
 {\fL  = 0.745^{+0.048}_{-0.058} \stat \pm 0.034  \syst}. 
\]

\section{Branching fraction determination}
\label{sec : BFresults}

The branching fraction of  the decay mode \BdtoRhoRho relative to the decay \BdtoPhiKst can be expressed as  

\begin{multline}
\label{equation : BFdefinitionFinal}
\frac{\BR(\BdtoRhoRho)}{\BR(\BdtoPhiKst)} = 
\frac{\lambda_{\fL} \cdot P(\BdtoRhoRho)}{P(\BdtoPhiKst)}
\times \frac{N'(\BtoPiPiPiPi)}{N'(\BtoKKKPi)} \\
\times \frac{\BR(\phi \to \kaon^+ \kaon^-)\BR(\kaon^* \to \kaon^+ \pi^-)}{\BR(\rho^0\to\pi^+\pi^-)^2}, 
\end{multline}

\noindent where the factor $\lambda_{\fL}$ corrects for differences in detection efficiencies between experimental and simulated data 
due to the polarisation hypothesis of the \BdtoRhoRho sample,
 $P(\BdtoRhoRho)$ and $P(\BdtoPhiKst)$ are the fractions of \BdtoRhoRho and \BdtoPhiKst signals in the samples of $\Bd \to (\pi^+\pi^-)(\pi^+\pi^-)$~and $\Bd \to (K^+K^-)(K^+\pi^-)$~decays, respectively. The quantities $N'(\BtoPiPiPiPi)$ and $N'(\BtoKKKPi)$ are the yields of \BtoPiPiPiPi~and $\Bd \to (K^+K^-)(K^+\pi^-)$~decays as determined from a fit to the four-body mass distributions, weighted for each data-taking period by the efficiencies of the signal and normalisation channels obtained from their respective simulated data. Finally, $\BR(\phi \to \kaon^+ \kaon^-)$, $\BR(\Kstarz \to \kaon^+ \pi^-)$ and $\BR(\rho^0 \to \pi^+\pi^-)$ denote known branching fractions~\cite{PDG2014}.

The product  $\lambda_{\fL} \cdot P(\BdtoRhoRho)$ is determined from the amplitude analysis to be ${1.13 \pm 0.19 \stat \pm 0.10 \syst}$.  This quantity is mainly related to the modelling of the \swave component, and dominates the systematic uncertainty of the parameters of interest.

The fraction of \BdtoPhiKst present in the $\Bd \to (K^+K^-)(K^+\pi^-)$ sample is taken from Ref.~\cite{LHCb-PAPER-2014-005}. 
A 1\% systematic uncertainty is added, accounting for differences in the selection acceptance for P- and S-wave contributions.

The amounts of \BtoPiPiPiPi and $\Bd \to (K^+K^-)(K^+\pi^-)$~candidates are determined from the four-body mass spectra analysis and their associated statistical and systematical uncertainties are propagated quadratically to the branching fraction uncertainty estimate.

The limited size of the simulated events samples that meet all selection criteria 
result in a systematic uncertainty of 1.7\% (2.6\%) 
on the measurement of the relative branching fraction for the 2011 (2012) data-taking period. The impact of the discrepancies between experimental and simulated data related to the \Bd meson kinematical properties is 0.6\% (1.2\%). The efficiencies of the particle-identification requirements are determined from control samples of data  
with a systematic uncertainty of 0.5\%, mostly originating from the limited size of the calibration samples. 
An additional 1\% systematic uncertainty on the tracking efficiency is added accounting for different interaction lenghts between \pion and \kaon.

The relative branching fraction is measured to be  

\begin{equation}
\dfrac{\BR(\BdtoRhoRho)}{\BR(\BdtoPhiKst)} = 0.094 \pm 0.017 \stat \pm 0.009 \syst . 
\label{equation : BFratio}
\end{equation}
\noindent
The agreement between the results obtained in the two data-taking periods is tested with the best linear estimator technique~\cite{Lyons:1988rp} yielding compatible results. 

The average branching fraction of \BdtoPhiKst as determined in Ref.~\cite{PDG2014} does not take into account the correlations between systematic uncertainties due to the S-wave modelling. Instead, we average the results from
Refs.~\cite{Aubert:2008zza,Prim:2013nmy,Briere:2001ue}
including these correlations to obtain $\BR(\BdtoPhiKst) = (1.00 \pm 0.04 \pm 0.05) \times 10^{-5}$. Using this value in~\eq{BFratio}, the branching fraction of \BdtoRhoRho is 
\begin{equation}
\BR(\BdtoRhoRho) = (0.94 \pm 0.17 \stat \pm 0.09 \syst \pm 0.06 \; ({\rm BF}))  \times 10^{-6}, 
\nonumber
\end{equation}

\noindent where the last uncertainty is due to the normalisation channel branching fraction.
Using the $\BdtoRhoRho$ branching fraction, the $\rho^0 f_0(980)$ amplitude, a phase space correction 
and assuming 100\% correlated uncertainties, 
an upper limit 
for the \Bd$\to\rho^0 f_0 (980)$ decay, 
at 90\% confidence level, is obtained
\begin{equation}
\BR(\Bd\to\rho^0 f_0 (980)) \times \BR(f_0 (980) \to \pi^+\pi^-) < 0.81 \times 10^{-6}. \nonumber \\
\end{equation}

\section{Conclusions}
\label{sec : conclusions}

The full data set collected by the \lhcb experiment in 2011 and 2012,
corresponding to an integrated luminosity of 3.0\invfb, is analysed to search for the \BdtoRhoRho decay.
A yield of $634\pm 28 \pm 8$ \BtoPiPiPiPi signal decays with $\pi^+\pi^-$ pairs in the 300--1100\mevcc mass range is obtained. An amplitude analysis is conducted to determine the contribution from \BdtoRhoRho decays. This decay mode is observed for the first time with a significance of 7.1 standard deviations. 
In the same $\pi^+\pi^-$ pairs mass range, $\Bs \to (\pi^+\pi^-)(\pi^+\pi^-)$ decays are also observed with a statistical significance of more than 10 standard deviations.

The  longitudinal polarisation fraction of the \BdtoRhoRho decay is measured to be  ${\fL = 0.745^{+0.048}_{-0.058} \stat \pm 0.034 \syst}$.  The measurement of the \BdtoRhoRho branching fraction reads 
\[
\BF(\BdtoRhoRho) = (0.94 \pm 0.17 \stat \pm 0.09 \syst \pm 0.06 \; ({\rm BF})) \times 10^{-6}, 
\]
where the last uncertainty is due to the normalisation channel. These results are the most precise to date
and will improve the precision of the determination of the CKM angle $\alpha$.

The measured longitudinal polarisation fraction is consistent with the measured value from~\babar~\cite{Aubert:2008au} while it differs by $2.3$ standard deviations from the value obtained by~\belle~\cite{Adachi:2012cz}.
The branching fraction measurement is in agreement with the values measured by both~\babar~\cite{Aubert:2008au} and~\belle~\cite{Adachi:2012cz} collaborations.

The evidence of the $\Bd \to \rho^0 f_0(980)$ decay mode reported by the~\belle collaboration~\cite{Adachi:2012cz} is not confirmed, 
and an upper limit at 90\% confidence level is established 
\[
\BR(\Bd\to\rho^0 f_0 (980)) \times \BR(f_0 (980) \to \pi^+\pi^-) < 0.81 \times 10^{-6}. \nonumber \\
\]

\section*{Acknowledgements}

\noindent We express our gratitude to our colleagues in the CERN
accelerator departments for the excellent performance of the LHC. We
thank the technical and administrative staff at the LHCb
institutes. We acknowledge support from CERN and from the national
agencies: CAPES, CNPq, FAPERJ and FINEP (Brazil); NSFC (China);
CNRS/IN2P3 (France); BMBF, DFG, HGF and MPG (Germany); INFN (Italy); 
FOM and NWO (The Netherlands); MNiSW and NCN (Poland); MEN/IFA (Romania); 
MinES and FANO (Russia); MinECo (Spain); SNSF and SER (Switzerland); 
NASU (Ukraine); STFC (United Kingdom); NSF (USA).
The Tier1 computing centres are supported by IN2P3 (France), KIT and BMBF 
(Germany), INFN (Italy), NWO and SURF (The Netherlands), PIC (Spain), GridPP 
(United Kingdom).
We are indebted to the communities behind the multiple open 
source software packages on which we depend. We are also thankful for the 
computing resources and the access to software R\&D tools provided by Yandex LLC (Russia).
Individual groups or members have received support from 
EPLANET, Marie Sk\l{}odowska-Curie Actions and ERC (European Union), 
Conseil g\'{e}n\'{e}ral de Haute-Savoie, Labex ENIGMASS and OCEVU, 
R\'{e}gion Auvergne (France), RFBR (Russia), XuntaGal and GENCAT (Spain), Royal Society and Royal
Commission for the Exhibition of 1851 (United Kingdom).

\addcontentsline{toc}{section}{References}
\setboolean{inbibliography}{true}
\bibliographystyle{LHCb}
\bibliography{main,LHCb-PAPER,LHCb-DP}

\newpage
\centerline{\large\bf LHCb collaboration}
\begin{flushleft}
\small
R.~Aaij$^{41}$, 
B.~Adeva$^{37}$, 
M.~Adinolfi$^{46}$, 
A.~Affolder$^{52}$, 
Z.~Ajaltouni$^{5}$, 
S.~Akar$^{6}$, 
J.~Albrecht$^{9}$, 
F.~Alessio$^{38}$, 
M.~Alexander$^{51}$, 
S.~Ali$^{41}$
% ...

{\footnotesize \it
% ...
$ ^{5}$Clermont Universit\'{e}, Universit\'{e} Blaise Pascal, CNRS/IN2P3, LPC, Clermont-Ferrand, France\\
$ ^{6}$CPPM, Aix-Marseille Universit\'{e}, CNRS/IN2P3, Marseille, France\\
% ...
$ ^{9}$Fakult\"{a}t Physik, Technische Universit\"{a}t Dortmund, Dortmund, Germany\\
% ...
$ ^{37}$Universidad de Santiago de Compostela, Santiago de Compostela, Spain\\
$ ^{38}$European Organization for Nuclear Research (CERN), Geneva, Switzerland\\
% ...
$ ^{41}$Nikhef National Institute for Subatomic Physics, Amsterdam, The Netherlands\\
% ...
$ ^{46}$H.H. Wills Physics Laboratory, University of Bristol, Bristol, United Kingdom\\
% ...
$ ^{51}$School of Physics and Astronomy, University of Glasgow, Glasgow, United Kingdom\\
$ ^{52}$Oliver Lodge Laboratory, University of Liverpool, Liverpool, United Kingdom\\
% ...
}
\end{flushleft}

\end{document}